\documentclass[%
 reprint,
showpacs,preprintnumbers,
 amsmath,amssymb,
 aps,
]{revtex4-1}
\usepackage{stmaryrd}
\usepackage{mathrsfs}
\usepackage{amssymb}
\usepackage{amsmath}

\usepackage{graphicx}
\usepackage{bm}

\begin{document}


\title{Growing Random Geometric Graph Models of Super-linear Scaling Law}

\author{Jiang Zhang}
 \email{zhangjiang@bnu.edu.cn}

\affiliation{Department of Systems Science, School of Management, Beijing Normal University}%

\date{\today}

\begin{abstract}
Recent researches on complex systems highlighted the so-called
super-linear growth phenomenon. As the system size $P$ measured as
population in cities or active users in online communities
increases, the total activities $X$ measured as GDP or number of new
patents, crimes in cities generated by these people also increases
but in a faster rate. This accelerating growth phenomenon can be
well described by a super-linear power law $X \propto
P^{\gamma}$($\gamma>1$). However, the explanation on this phenomenon
is still lack. In this paper, we propose a modeling framework called
growing random geometric models to explain the super-linear
relationship. A growing network is constructed on an abstract
geometric space. The new coming node can only survive if it just
locates on an appropriate place in the space where other nodes
exist, then new edges are connected with the adjacent nodes whose
number is determined by the density of existing nodes. Thus the
total number of edges can grow with the number of nodes in a faster
speed exactly following the super-linear power law. The models
cannot only reproduce a lot of observed phenomena in complex
networks, e.g., scale-free degree distribution and asymptotically
size-invariant clustering coefficient, but also resemble the known
patterns of cities, such as fractal growing, area-population and
diversity-population scaling relations, etc. Strikingly, only one
important parameter, the dimension of the geometric space, can
really influence the super-linear growth exponent $\gamma$.

\end{abstract}


\pacs{89.75.-k,89.75.Da}
\maketitle


\section{Introduction}
\label{sec.introduction} The super-linear phenomenon is described as
a scaling relation,
\begin{equation}
\label{eq.superlinearlaw} X=c P^{\gamma}.
\end{equation}
$X$ and $P$ may have different representations in different systems.
In urban systems, for example, $X$ represents GDP, R\&D investments,
crimes or the number of new patents, and $P$ represents the
population\cite{bettencourt_growth_2007,bettencourt_invention_2007,bettencourt_unified_2010,bettencourt_urban_2010}.
In online communities, $X$ is the total number of activities (tags,
blogs) generated by the users, and $P$ is the total number of active
users (who at least generate one
activity)\cite{wu_accelerating_2011}. In language, $X$ is the total
number of words in an article, $P$ is the number of distinct words
in the same article\cite{leijenhorst_formal_2005,lu_zipfs_2010}. In
equation \ref{eq.superlinearlaw}, $\gamma$ is an exponent to
describe the relative speed of $X$ respective to $P$. A large number
of empirical studies reported that $\gamma$ is always falling into
the interval $[0,2)$. For example, \cite{wu_accelerating_2011}
pointed out $\gamma$s are $1.17\sim 1.48$ for different online
communities. \cite{bettencourt_growth_2007} finds $\gamma$s are
$1.15\sim 1.26$ for cities in different countries. However, the
exponent for the relationship of population and GDP can approach to
$1$ if the scale of the system is large and interactions among
people are weak. For example, \cite{zhang_allometric_2010} found
that the exponent is almost $1$ for countries. According to our
unpublished results, the scaling relationship is almost linear for
provinces and states. So far, we know the equation
\ref{eq.superlinearlaw} holds for a large number of different
systems, but the exponents are always different system by system.
While, the next question is what is the underlying mechanism of this
remarkable phenomenon?

There are already some studies trying to explain the super-linear
growth phenomenon. For instance, Arbesman et al. tried to attribute
the super-linear phenomenon to the properties of the interaction
network\cite{arbesman_superlinear_2009}, but their model takes
several assumptions on the network which are hardly to find the
correspondence in the real systems. While
\cite{wu_accelerating_2011,lu_zipfs_2010} tried to link the
universal patterns in distributions (e.g. DGBD distribution in
\cite{wu_accelerating_2011} and Zipf law in \cite{lu_zipfs_2010}) to
the super-linear growth pattern by large number of empirical data.
Despite a strong connection between size-dependent distributions and
super-linear growth is revealed \cite{wu_accelerating_2011}, the
underlying mechanisms are still unknown since size-dependent
distribution and super-linear growth actually are the two different
expressions for the same law\cite{wu_accelerating_2011}.

In the network community, researchers have found many empirical
networks are of a so-called accelerating growth
phenomenon\cite{dorogovtsev_accelerated_2002,dorogovtsev_evolution_2001}
which also states the power law relationship between the number of
edges ($X$) and the number of nodes ($P$), but they didn't try to
explain this fact. Leskovec et al. \cite{leskovec_graphs_2005}
re-found the accelerating growth pattern and re-name it as the
densification phenomenon. He tried to build a forest fire model to
understand its origin\cite{leskovec_graphs_2005}. But due to the
complexity of this model, he later developed a totaly new one called
kronecker graph model. As claimed in \cite{leskovec_kronecker_2010},
densification phenomenon is a mathematical property of kronecker
products. Although it succeeds to fit many empirical network data,
the explanations and real life grounding are still lack. Also, in
kronecker graph model, the intercept of the power law relation
between number of nodes and edges, i.e., $c$ in equation
\ref{eq.superlinearlaw} must be 1. This strong assumption is hardly
supported by empirical data. However, these studies make us clear
that the super-linear growth pattern widely existing in various
systems can be discussed on a network background. Recently, by
analyzing the data of cell-phone communication networks in different
cities, Schlapfer et al.\cite{schlapfer_scaling_2012} found the
accelerating growth exponent $\gamma$ is of the same value as the
super-linear growth exponent of cities and the clustering
coefficients in these networks are size invariant. This coefficient
almost determines the super-linear growth
exponents\cite{schlapfer_scaling_2012}. Therefore, as the size of
the network increases, the clustering coefficient must keep
unchanged so that the accelerating growth or densification pattern
as a systemic results can emerge. However, their model cannot answer
what is the origin of the size-invariant clustering coefficient, so
the super-linear growth puzzle also remained unsolved. More
recently, Bettencourt developed a network model to explain the
origin of the super-linear growth in urban
systems\cite{bettencourt_origins_2012}. Although this model can fit
the empirical data of cities very exactly, it is complicated and
depends on a set of assumptions which are hardly tested.

Despite several models have been presented to explain the
super-linear growth scaling law, we still cannot find one simple
model with minimum parameters while can reproduce as many as
possible patterns observed in empirical systems. In this paper, we
propose a new growing network modeling framework in geometric space
called growing random geometric models to explain the super-linear
phenomenon. It uses very basic but simple mechanism to reproduce a
lot of observed patterns in cities and networks. Strikingly, we
found the super-linear exponent is determined only by one important
parameter, $d$, the dimension of the geometric space.

\section{Basic Model}
\label{sec.basicmodel}

Inspired by the niche model in food web
studies\cite{williams_simple_2000}, we can construct a spatial
growing network in an abstract geometric space. If the new coming
node just locate on the right place which can match existing nodes,
then the new one can survive and some new links are built
accordingly.

This basic idea is very similar to the well developed model called
random geometric graph\cite{penrose_random_2003} and disk
percolation\cite{newman_fast_2001}, the main difference is the
growing mechanism in our model. Unlike some well known growing
network
models\cite{barabasi_emergence_1999,dorogovtsev_evolution_2001}, the
number of new coming edges is not given but determined by the
existing nodes. We will introduce one of the simplest model of this
framework in this section and left more interesting extensions to
the following sections.

The basic model contains following elements: a geometric space
$\mathscr{S}$ which can be modeled as a $d$ dimensional Euclidean
space, in which the coordinates can be any real numbers, that is,
$\mathscr{S}=\mathscr{R}^d$, where $\mathscr{R}$ is the set of real
numbers. A relation as the matching rule $R$ is defined on
$\mathscr{S}$, $R\in \mathscr{S}\times \mathscr{S}$. In the basic
model, we can set the simplest matching rule as the Euclidean
distance between two points cannot be exceed a given parameter $r$,
that is:
\begin{equation} \label{eq.matchingrule1}
R=\{(x,y)\in \mathscr{S}\times \mathscr{S}| \bigparallel
x-y\bigparallel <r\}
\end{equation}

The simplest initial condition is the geometric space contains only
a single agent locates in the origin $0\in \mathscr{S}$. We of
course can design more complicated initial conditions in the
extended model.

The growing process of this model is like this.

{Step 1}: In each time step, one new agent $i$ is added in the
system with a randomly assigned coordinate $x_i\in \mathscr{S}$, if
some existing agents' coordinates match the new one, then $i$ may
survive, otherwise it may die immediately. We denote the set of
existing agents who have matched with agent $i$ as
$M_i=\{j|\bigparallel x_i-x_j\bigparallel <r\}$, then new coming
agent $i$ can only survive and exist in the system (keeps its
coordinate fixed) forever if $M_i\neq \O $.

\begin{figure}
\includegraphics[scale=0.6]{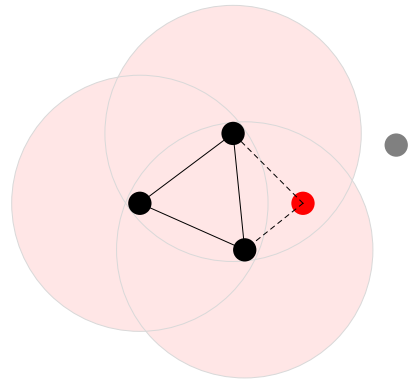}
\caption{A 2-d Geometric Space of the Basic Model. Black disks are
existing agents, the red disk is a new coming agent who will survive
while the gray one is the agent who cannot survive. The dark lines
are links between agents, the dashed lines are the adding links
between the red agent and existing agents.} \label{fig.2dexample}
\end{figure}

{Step 2}: If the new coming agent survives, then new links are added
from the agents in $M_i$ to the new one $i$ (As shown in figure
\ref{fig.2dexample}).

Then, we will repeat these two steps to obtain a growing network.
Through studying the relationship between the total number of edges
$X$ and the number of nodes in the network, $P$, we can test the
super-linear growth law.

Although the growing process is very simple and seems homogenous,
the resulting network is very uneven both in time and space. First,
the new agent is added in the system in a random place of the
geometric space, but only if the random place is surrounded by
existing agents, the new place will be occupied. Thus the density of
existing agents is uneven in the geometric space, the network itself
can be regarded as a result of ``crystalization''. Second, the
growing process is uneven in time because a much slower growing
speed is expected in the initial process than the following steps.

However, in the simulation, we have to use a trick to avoid the
problem of random searching on an infinite space: a new coming
agent's coordinate $x_i$ is not randomly assigned in the whole
geometric space $\mathscr{S}$ but a much smaller subset
$\mathscr{T}=\{y|y\in [\eta x_m,\eta x_M]\}$, where $\eta=5$ in the
simulations, $x_m,x_M$ are minimum and maximum coordinates along all
dimensions. In a word, the new coming agent is from a
$d$-dimensional box covering all existing agents randomly. This
trick can accelerate our program dramatically but take no effect on
the final results.

\subsection{One Dimensional Model}
Let's consider the simplest case of our basic model, the geometric
space is a one dimensional line, i.e., $d=1$. In this simplest case,
the super-linear growth phenomenon can be generated.
\begin{figure}
\includegraphics[width=9cm]{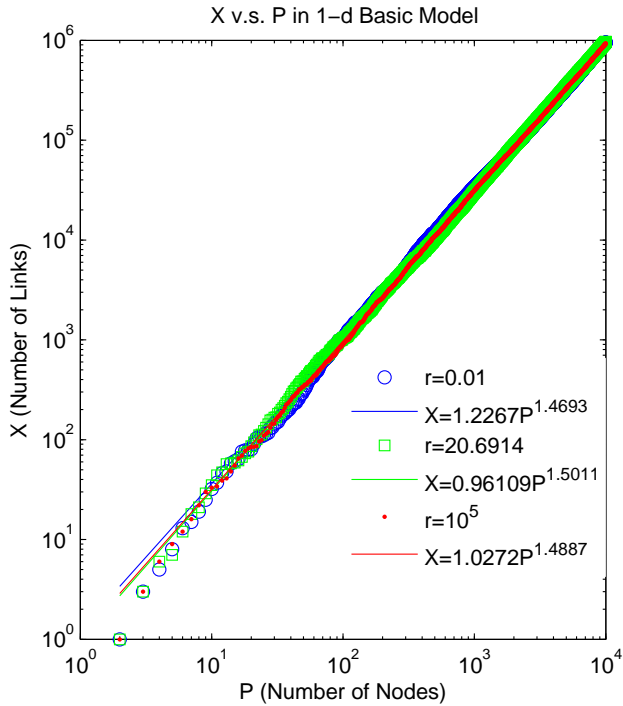}
\caption{Number of Edges v.s. Number of Nodes in 1-d Basic Model
with Different $r$} \label{fig.1dpopgdp}
\end{figure}

\begin{figure}
\includegraphics[scale=1]{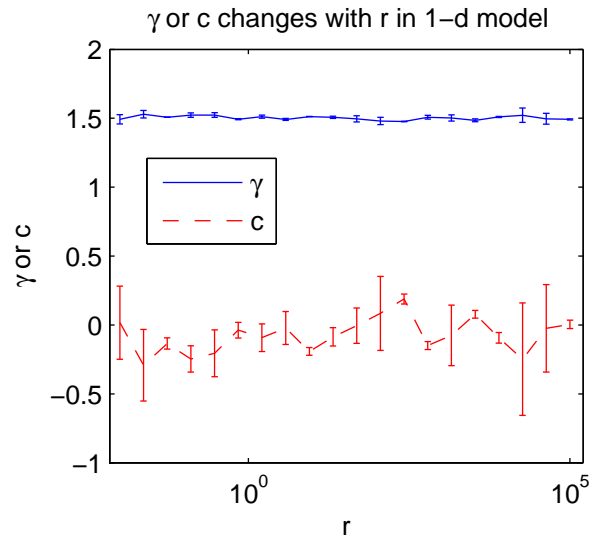}
\caption{Parameters $\gamma,c$ change with $r$, all the simulations
are done by 10 times with 10000 time steps.}
\label{fig.1dparameters}
\end{figure}

Figure \ref{fig.1dpopgdp} shows three simulations with different
$r$. We found at first all the simulations show super-linear growth,
that is the number of edges v.s. the number of nodes in different
time step has a power law relation with exponent larger than one.
Second, all the curves of $X$ v.s. $P$ almost overlap each other on
the plot which means the fittings by equation
\ref{eq.superlinearlaw} have nearly same parameters. So, the
exponents in equation \ref{eq.superlinearlaw} are independent on the
parameter $r$. This point can be confirmed by the larger scale
simulations as shown in \ref{fig.1dparameters}.

We observed clearly both $\gamma$ and $c$ fluctuate around the mean
values in different $r$. Therefore, the super-linear growth
phenomenon doesn't dependent on the parameter $r$.

\subsection{Two Dimensional Model}
Besides the basic phenomena shown in the 1-d model, 2-d model shows
more interesting patterns. In this case, the geometric space itself
can be illustrated by a 2-d picture. And the network formed by the
model is a spatial network, so we can show the networks in different
steps.

\begin{figure}
\includegraphics[scale=0.8]{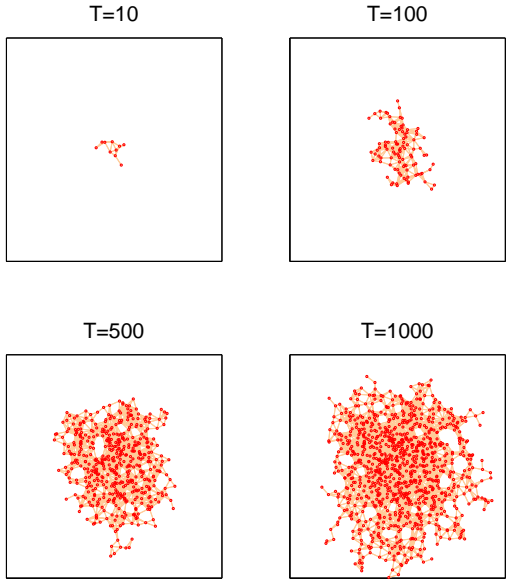}
\caption{Network formation in 2-d basic model ($r=10^5$) of
different time steps.} \label{fig.spatialnetworks}
\end{figure}

From figure \ref{fig.spatialnetworks}, we know that the growing
network in the geometric space is very uneven. We found the density
of agents in the center of the geometric space is much higher than
the peripheral places. Actually, the network in the 2-d geometric
space is a fractal. That is the number of occupied lattice scales as
the measurement size with the power law exponent (fractal dimension)
$\alpha$. This point can be confirmed by the box-counting method,
and the fractal dimensions is calculated in different time steps.
\begin{figure}
\includegraphics[scale=0.7]{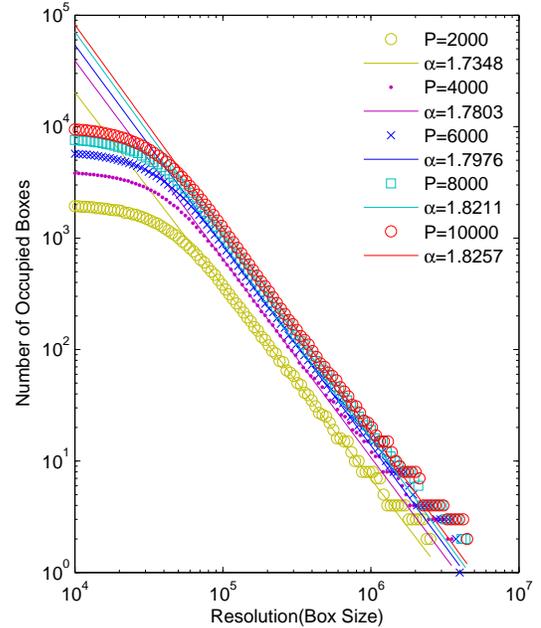}
\caption{Box counting calculation of fractal dimension $\alpha$ of
the growing network in different time steps ($r=10^5$).}
\label{fig.fractaldimensions}
\end{figure}

According to the box-counting method, we know the asymptotic fractal
dimension is about $1.83$. All the dimensions $\alpha$ during the
simulation are in between 1 and 2, therefore, the spatial networks
are fractals.

\subsection{Three Dimension Model}
So far all the geometric spaces we have discussed are very abstract.
In this subsection we will discuss a more concrete model: an
interaction network of a city. Each node on the network is an
individual living in the city, and the links between the nodes stand
for the interactions (e.g. phone connection or friendship
connection). The geometric space is a 3 dimensional Euclidean space
in which two dimensions stand for the geographic space (since a city
locate on a 2-dimensional plane of course) and the left single
dimension is the similarity space. The basic matching rules are the
same as the previous model settings. Hence, a connection is built
only if two individuals locate very closed in the geographic space
and have common interests (similarities).

In this model, all the phenomena we have discussed in last sections
can be also observed. For example, the super-linear exponent is
about $1.218$, the fractal dimension of the network in 3-d space is
about $2.36$. While the projection of the network on the geographic
space (2-d world) is also a fractal with dimension around $1.77$
being smaller than the one's in the 2-d model. Thus the complexity
of the 2-d projection of the 3-d network is smaller than the 2-d
model because the new introducing dimension of the model make the
matching criterion stricter.

\begin{figure}
\includegraphics[scale=0.8]{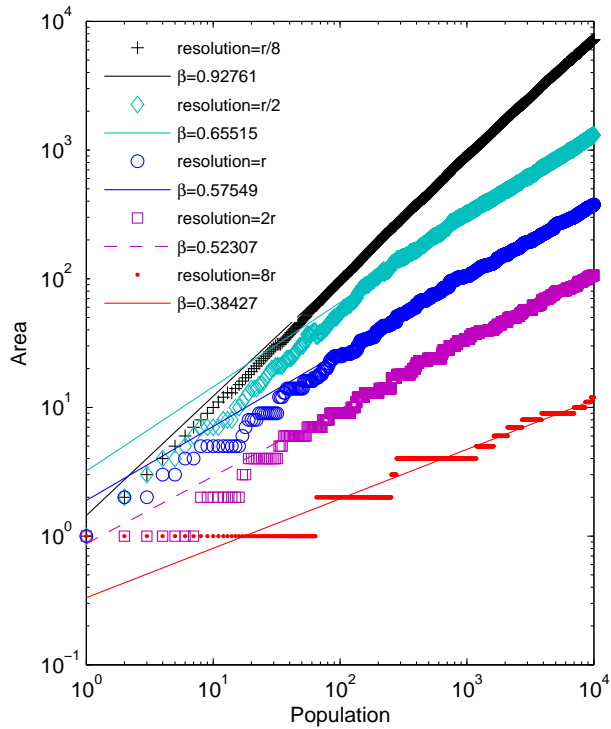}
\caption{Area-population relationship in different resolutions.
Here, resolution is the size of the box to cover the set of nodes in
the 2-d space, area is the number of boxes counted by the given
resolution} \label{fig.areapop}
\end{figure}

Besides the fractal dimension of the network, we can also study the
relationship between area and population which is comparable to the
empirical studies in
cities\cite{nordbeck_urban_1971,lee_allometric_1989}. In our model,
we calculate city's area by the following method. On the 2-d
geographic space, we select a specific resolution as our
observational scale. Then we use the given resolution to rasterize
the whole geographic space, after that, we count the number of
occupied boxes as the area just like the box-counting method in the
fractal dimension calculation. Because the box occupied by multiple
agents is treated as one unit of area, the increase speed of area is
much slower than the speed of population increasing. Therefore, a
sub-linear area-population relationship can be obtained as shown in
figure \ref{fig.areapop}

The area and population has a sub-linear power law relation:
\begin{equation} \label{eq.areapop}
A\propto P^\beta,
\end{equation}

Where, $A$ is the area of cities, and $\beta$ is the exponent. As
shown in figure \ref{fig.areapop}, all $\beta$s are smaller than 1
and fall into the interval $[0.38~0.97]$. This result is consistent
with the observed exponents $[0.33~0.91]$ of real
cities\cite{nordbeck_urban_1971,lee_allometric_1989}. We also show
how the area-population relation depend on resolution. As the size
of the box increases, $\beta$ increases also. Because city is a
fractal object, the area as a macro measurement is dependent on the
measurement scale certainly.

\begin{figure}
\includegraphics[scale=0.8]{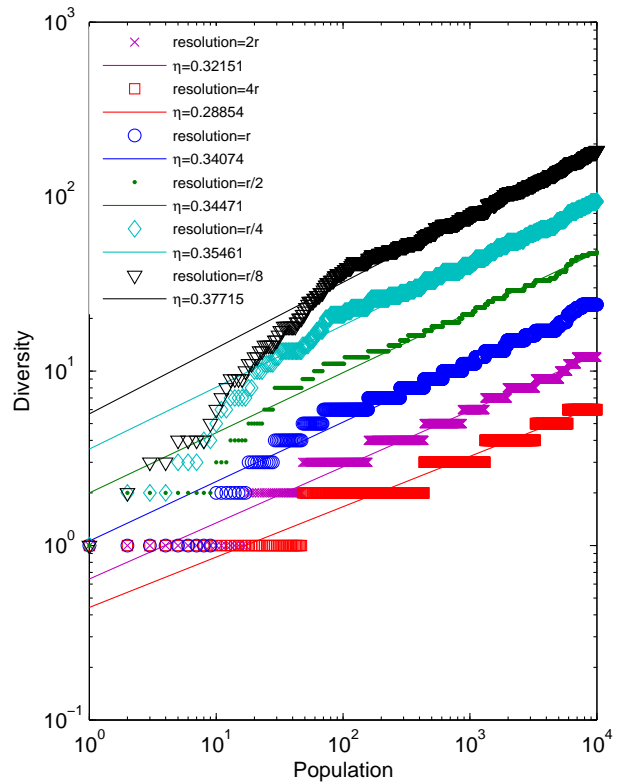}
\caption{Diversity-population relationship in different area
resolution.} \label{fig.diversitypop}
\end{figure}

We can use the similar method to study the similarity space and
found similar sub-linear law between diversity (different types of
features) and population,
\begin{equation} \label{eq.divpop}
D\propto P^\eta,
\end{equation}
In figure \ref{fig.diversitypop}, all the exponents $\eta$s are
around 0.3 which are much smaller than $\beta$ and more stable with
respect to different resolutions.

Beyond the spatial properties and super-linear growth, we can also
discuss other network features, and how do they change with the size
of the system.
\begin{figure}
\includegraphics[scale=0.8]{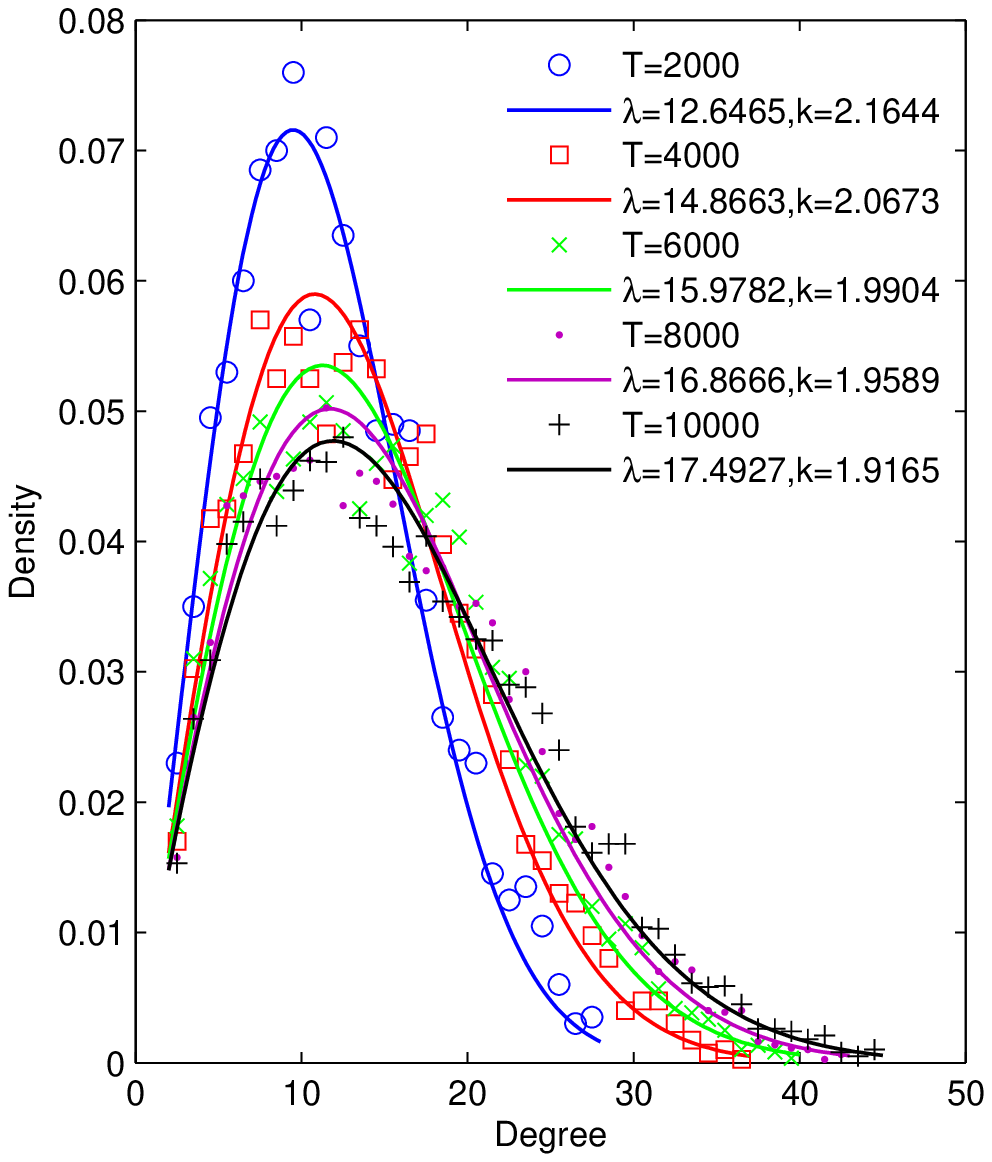}
\caption{Degree Distributions with Time. We use weibull distribution
($p(x)=\frac{k}{\lambda}(\frac{x}{\lambda})^{k-1}e^{-(\frac{x}{\lambda})^k}$)
curve to fit the simulation data.} \label{fig.degreedistribution}
\end{figure}
\begin{figure}
\includegraphics[scale=0.7]{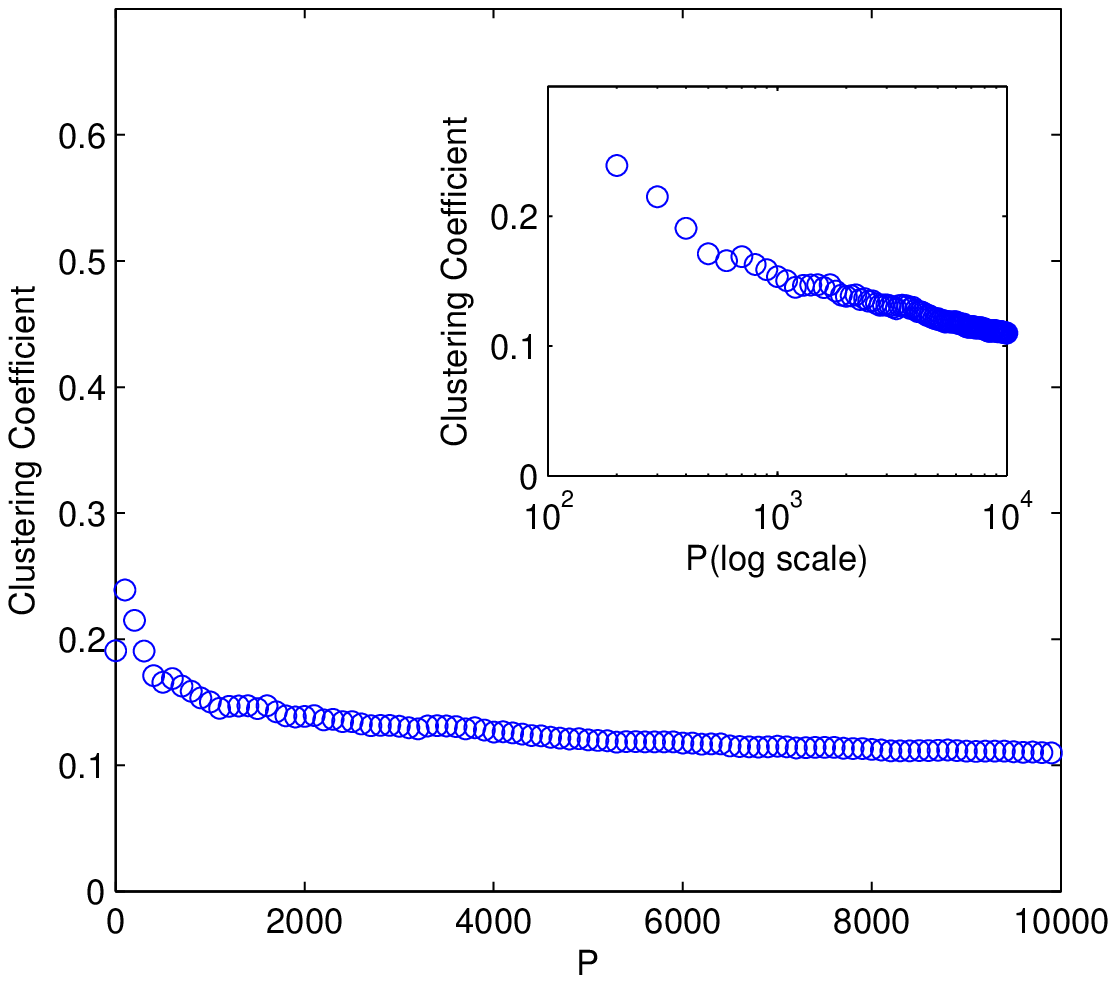}
\caption{Clustering coefficient change with size $P$}
\label{fig.clusteringc}
\end{figure}

The degree distributions are not power laws but Weibull
distributions. This is inconsistent with the empirical observation
that the degree distributions are heavy tails. However, the
clustering coefficient asymptotically unchange with the size of the
network. This is also observed in empirical
data\cite{schlapfer_scaling_2012}.

\begin{figure}
\includegraphics[scale=0.8]{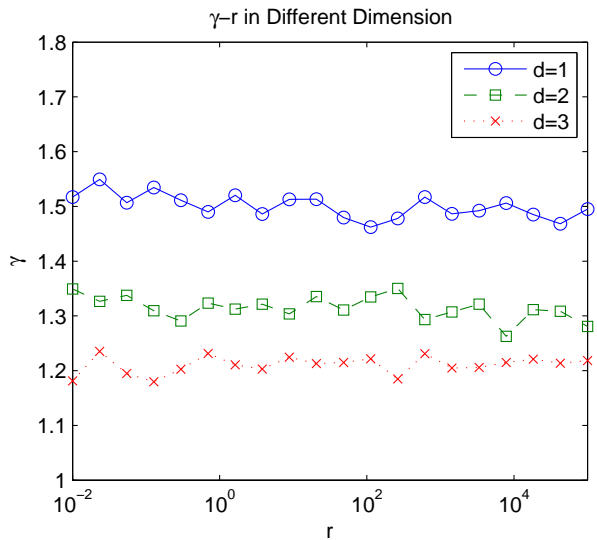}
\caption{Dependence of $\gamma$ on $r$ in different dimensions.}
\label{fig.gammar}
\end{figure}

\begin{figure}
\includegraphics[scale=0.8]{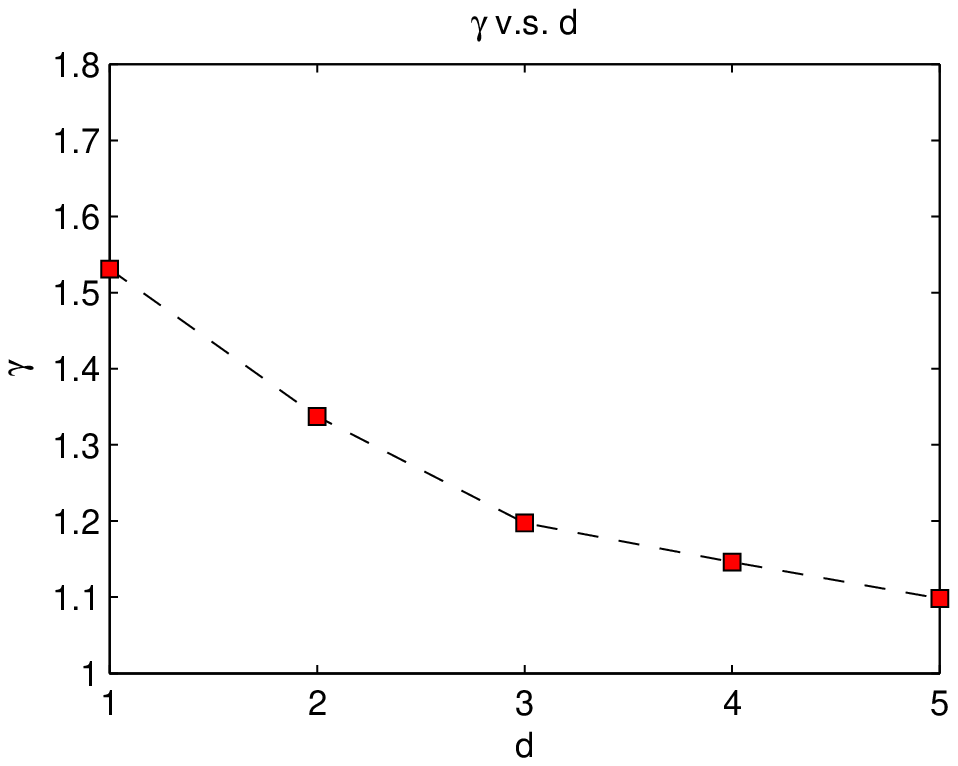}
\caption{Decay of $\gamma$ with the spatial dimension $d$.}
\label{fig.gammad}
\end{figure}

Interestingly, through the simulations in 1,2 and 3 dimensions, we
found the super-linear growth exponents depend not on $r$ but the
spatial dimension $d$ (as shown in figure \ref{fig.gammar}). To see
how does super-linear growth exponent decay with the spatial
dimension, we have done more experiments as shown in figure
\ref{fig.gammad}.

\section{Model Extensions}
We have known what diverse and interesting patterns can the basic
model exhibit, however, we can add a little of complexity on the
basic one to make it closer to the reality. We will mainly consider
several possible extensions. Firstly, we can study the geometric
space with limitations. Secondly, we can add more heterogeneity in
our model.

\subsection{Finite Geometric Space}
\begin{figure}
\includegraphics[scale=0.8]{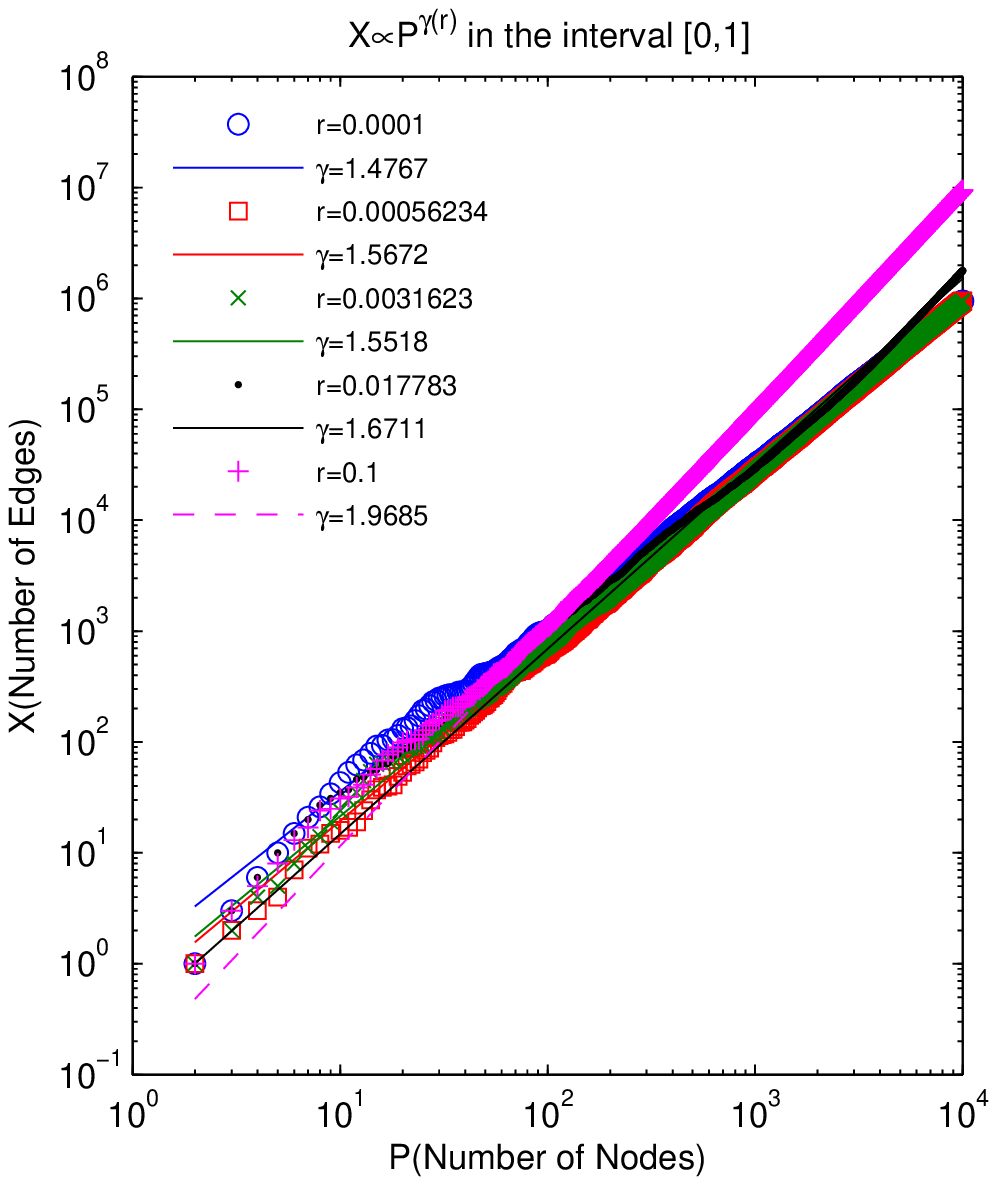}
\caption{$X\propto P^{\gamma}$ in Unit Interval}
\label{fig.finitespace}
\end{figure}

We will firstly extend our model to a finite geometric space. The
simplest finite one that we can imagine is the unit interval $[0,1]$
on the real line. In this case, the super-linear exponent will
depend on the interaction radius $r$ because the space is not
scale-free anymore but have a maximum characteristic scale which is
the upper bound of the radius. So, figure \ref{fig.finitespace}
shows different straight lines with different $r$.

As we observed, the slope of the straight line, i.e. the
super-linear exponent increases with the interaction radius $r$.
Even when the radius is large enough so that the scale is comparable
with the maximum range of the geometric space itself, then the power
law exponent $\gamma$ is approaching the maximum possible value $2$.

The unevenness of time, i.e., the waiting time between two agents
adding in the network, can be investigated in this extended model.
In section \ref{sec.basicmodel}, we have mentioned that a trick has
to be used to accelerate the simulation process otherwise infinite
time should be waited to add a new node when the geometric space is
infinite. However, in this extension, we can directly simulate the
whole random searching process without using this trick. In every
time step, a new agent with a random position in the interval is
added and survive with the condition that some old agents are close
to him. Therefore, as time goes by, the growing speed of the whole
network will be accelerated since the number of existing nodes
become larger and larger. Instead plotting the waiting time between
any two survival agents, we study the cumulative time, i.e. the
total time elapse $t$ so far versus the total number of agents
survived before $t$. We found a asymptotic power law between these
two variables.

\begin{equation}
\label{eq.timepowerlaw} t\propto P^{\chi}
\end{equation}

\begin{figure}
\includegraphics[scale=0.8]{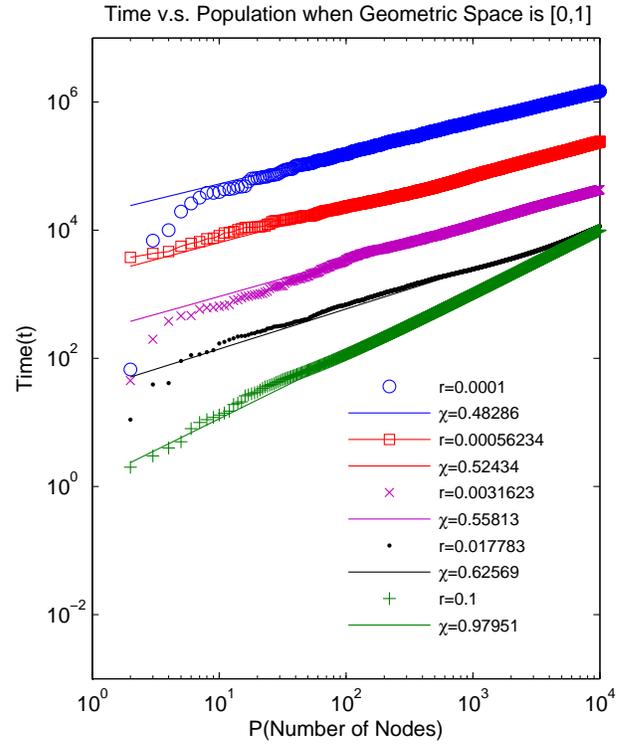}
\caption{The scaling relation between population $P$ with simulation
time $t$, $t\propto P^{\chi}$ in different $r$}
\label{fig.timefinite}
\end{figure}

From figure \ref{fig.timefinite}, we know the time intervals between
two agents added into the whole network scales with the size of the
network. And the exponent decreases with the interaction radius $r$.
When the radius $r$ is comparable with the scale of the geometric
space, the exponent $\chi$ is approaching $1$. Therefore, the
growing process is actually a fractional dynamical process.
\subsection{Finite Resolution}
In the previous subsection, we have considered the upper bound of
the size of the geometric space, the lower bound will be considered
in this subsection.

At first, we can model the whole geometric space as a discrete
cellular space. Each agent can only occupy one single cell. So the
new node can only exist only if (1) they can build a link with at
least one existing agent; (2) the new agent's position in the
geometric space is not occupied by any existing agent. By adding
this new rule, we find the super-linear exponent is dependent on the
interaction radius $r$.

\begin{figure}
\includegraphics[scale=0.8]{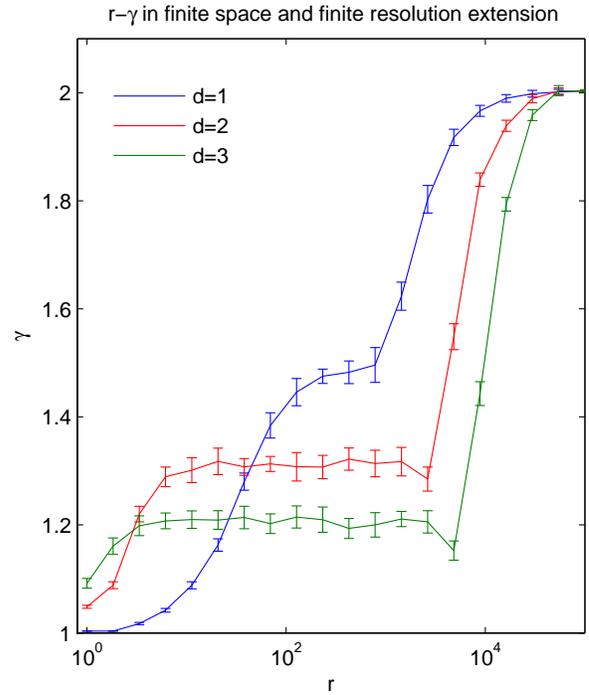}
\caption{The dependence of super-linear exponents on the interaction
radius $r$ in both $d=1,2,3$ dimensional spaces. All simulations are
run 20 times with $10^4$ cycles} \label{fig.finitelimitation}
\end{figure}

We set the minimum resolution as 1, and the maximum range of the
geometric space as $10^5$ in all the following simulations. The
interaction radius $r$ changes from $1$ to $10^4$, the dependence of
super-linear exponents on the radius $r$ in all $d=1,2,3$ space is
shown in figure \ref{fig.finitelimitation}

We see in all cases when $r=1$, the exponents are close to 1, which
means the networks are very regular and like lattices. As the
interaction radius $r$ increases, this constraint becomes weak, so
the exponents will increase also. When $r$ is in the intermediate
stage, the exponent $\gamma$s are always independent on $r$ because
both the upper limitation and lower limitation have no any
constraints on the systemic processes. The stationary exponents are
almost identical to the ones in the free geometric cases. When $r$
is big enough, the upper bound of the geometric space will influence
the behaviors of the $\gamma$s, so the exponents increase with the
interaction radius $r$ to reach the maximum value $2$. In these
extensions, we know the parameter $r$ can affect the super-linear
exponent $\gamma$ due to the space limitation effect.

\subsection{Heterogenous Models}
We found the degree distributions of the basic model are not power
laws as showed in many empirical networks. The essential reason is
the homogeneity of the basic model, i.e. all the interaction
radiuses are the same. This strong assumption is not supported by
real life. Thus, in this subsection, we will consider a heterogenous
model with random interaction radius.

In the first attempting, we suppose $r$ of each agent is a random
number following exponential distribution, so the cumulative
function of this variable is,

\begin{equation} \label{eq.exponentialdistribution}
Pr\{r\geq x\geq 0\}=\int_x^{+\infty}{\lambda \exp(-\lambda y)dy}.
\end{equation}

\begin{figure}
\includegraphics[scale=0.8]{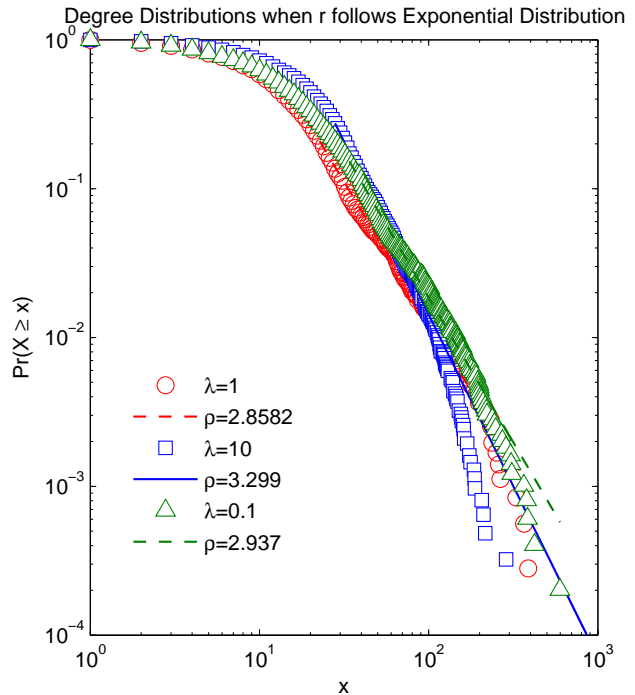}
\caption{Degree distributions in heterogenous model in 2-d
simulation with $r$ exponentially distributed in different
$\lambda$s. $\rho$s are the power law exponents of the degree
distributions} \label{fig.distributionexponential}
\end{figure}

In this way, we can generate both super-linear growth and scale-free
degree distribution patterns. That is the resulted degree
distribution has a power law tail.

\begin{equation} \label{eq.powerdistribution}
Pr\{k\geq x\geq m\}=\int_x^{+\infty}{(\rho-1)m^{\rho-1}y^{-\rho}dy},
\end{equation}
where, $k$ is the random variable for degrees, $m$ is the lower
degree of power law tail, $\rho$ is its exponent. As $\rho$
increases, the heterogeneity of the degrees becomes larger. Figure
\ref{fig.distributionexponential} shows the cumulative degree
distributions of several networks with different $\lambda$ values.

Exponential distribution of $r$ is not the only choice, we can use
other distribution density function to reproduce the power law
degree distribution and super-linear growth pattern. For example, we
replace the formula \ref{eq.exponentialdistribution} to:
\begin{equation} \label{eq.normaldistribution}
Pr\{r\geq x\geq
0\}=\int_x^{+\infty}{\frac{\sqrt{2}}{\sigma\sqrt{\pi}}
\exp(\frac{-y^2}{2\sigma^2})dy},
\end{equation}

\begin{figure}
\includegraphics[scale=0.8]{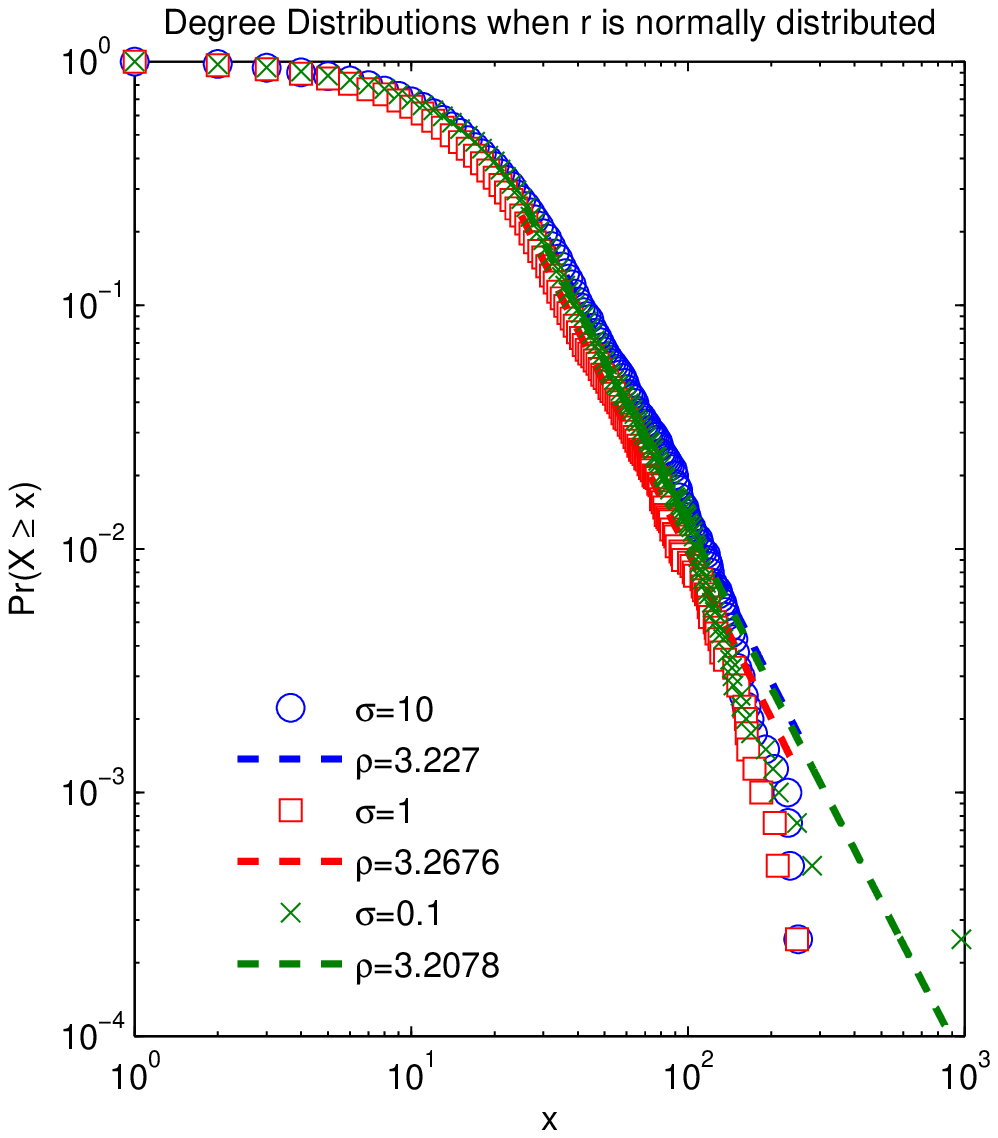}
\caption{Degree distributions in heterogenous model in 2-d
simulation with $r$ exponentially distribute and different $\sigma$.
$\rho$s are the power law exponents of degree distributions}
\label{fig.distributionnormal}
\end{figure}

That means $r$ follows the half normal distribution (in each time
step, draw a random number with normal distribution and take the
absolute value). Figure \ref{fig.distributionnormal} shows the
degree distributions.

Comparing figure \ref{fig.distributionexponential} and
\ref{fig.distributionnormal} we know the exponent of the degree
distribution in normal distribution model is larger than the one in
exponential distribution model. That means the heavy tail phenomenon
is more insignificant than the former case.

\begin{figure}
\includegraphics[scale=0.8]{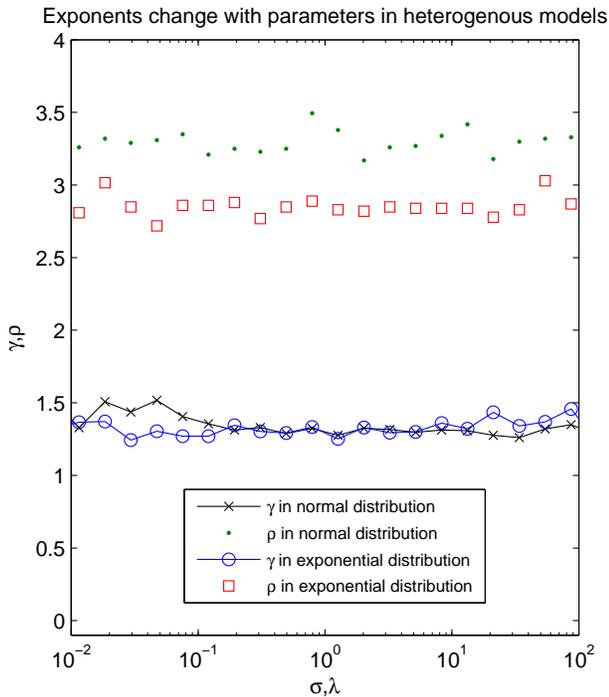}
\caption{How exponents $\gamma$ (allometry) and $\rho$ (scale-free)
change with parameters $\lambda$ (exponential distribution) and
$\sigma$ (normal distribution) in 2-d space}
\label{fig.allexponents}
\end{figure}

To see how do the exponents $\gamma$ and $\rho$ change with the
parameters $\lambda$ and $\sigma$, we have conducted larger scale
experiments. The results are shown in figure \ref{fig.allexponents}.
Both exponents $\gamma$ and $\rho$ are almost invariant when
$\lambda$ and $\sigma$ change. And because all the experiments are
done in 2-d space, the exponent $\gamma$ is almost identical to the
values in the basic model. Therefore, although we have to introduce
two new parameters $\lambda$ and $\sigma$, the super-linear exponent
is independent on them.

\section{Discussion}
In this paper, we introduce a new growing network model called
growing random geometric graph. Actually, this is a modelling
framework that can be used to model various complex networks and
other systems. One of the main advantages of these models is they
all exhibit super-linear growth or densification, accelerating
growth phenomenon.

Besides the super-linear growth behavior, this simple model can also
show a lot of scaling behaviors. We used a set of exponents to
characterize these scalings. $\alpha$ is the fractal dimension of
the spatial network in the space, $\beta$ is the exponent of area
and population, $\eta$ characterize the scaling between diversity of
similarities and population, $\chi$ describe the power law relation
between time and the size of the system and $\rho$ is the power law
exponent of the degree distribution in the extended model. All these
scaling behaviors indicate that the growing random geometric graph
is an anomalous object that is governed by some unknown fractional
dynamics. Further studies, especially the mathematical analysis are
deserved.

Although we have discussed several interesting extensions toward the
original model, more extensions are needed. For example, we can grow
the network not only in the Euclidean space but other interesting
space, e.g. hyperbolic space\cite{papadopoulos_popularity_2012}. And
other possible matching rules can be considered. Maybe more
interesting phenomena will emerge.

Finally, this is only the first step of this model, both the
theoretical analysis and empirical tests are needed in the future
studies.

\begin{acknowledgements}
Thanks for the discussions with Prof. Bettencourt in Santa Fe
Institute, doctor Wu in Hong Kong city university and Prof. Wang and
Chen in beijing normal university, acknowledges the support from the
National Natural Science Foundation of China under Grant No.
61004107.
\end{acknowledgements}

\bibliography{references}

\end{document}